\documentclass{emulateapj}
\usepackage{lscape}

\def\e{{\rm E}}

\def\max{{\rm max}}

\def\e{{\rm E}}

\begin{document}
\title{A Jovian-mass Planet in Microlensing Event OGLE-2005-BLG-071} 

\author{A.~Udalski\altaffilmark{1},
M.~Jaroszy{\'n}ski\altaffilmark{1}, 
B.~Paczy{\'n}ski\altaffilmark{2},
M.~Kubiak\altaffilmark{1},
M.K.~Szyma{\'n}ski\altaffilmark{1},
I.~Soszy{\'n}ski\altaffilmark{1,3},
G.~Pietrzy{\'n}ski\altaffilmark{1,3},
K.~Ulaczyk\altaffilmark{1},
O.~Szewczyk\altaffilmark{1},
{\L}.~Wyrzykowski\altaffilmark{1,4}\\
(The OGLE Collaboration), \\
G.W. Christie\altaffilmark{5},
D.L. DePoy\altaffilmark{6},
S. Dong\altaffilmark{6},
A. Gal-Yam\altaffilmark{7},
B.S. Gaudi\altaffilmark{8},
A. Gould\altaffilmark{6},
C. Han\altaffilmark{9},
S. L\'epine\altaffilmark{10},
J.~McCormick\altaffilmark{11},
B.-G. Park\altaffilmark{12},
R.W. Pogge\altaffilmark{6} \\
(The $\mu$FUN Collaboration), \\
D.P. Bennett\altaffilmark{13},
I.A. Bond\altaffilmark{14},
Y. Muraki\altaffilmark{15},
P.J. Tristram\altaffilmark{16},
P.C.M.~Yock\altaffilmark{16}\\
(From the MOA Collaboration)\\ 
J.P. Beaulieu\altaffilmark{17},
D.M. Bramich\altaffilmark{18,19},
S.W. Dieters\altaffilmark{20},
J. Greenhill\altaffilmark{20},
K. Hill\altaffilmark{20},
K. Horne\altaffilmark{18},
D. Kubas\altaffilmark{21}\\
(From the PLANET/RoboNet Collaboration)\\ 
\vspace{\baselineskip}
}

\altaffiltext{1}
{Warsaw University Observatory, Al.~Ujazdowskie~4, 00-478~Warszawa, Poland;
udalski@astrouw.edu.pl}
\altaffiltext{2}{Princeton University Observatory, Princeton, NJ 08544, 
USA}
\altaffiltext{3}{Universidad de Concepci{\'o}n, Departamento de Fisica,
Casilla 160--C, Concepci{\'o}n, Chile}
\altaffiltext{4}
{Jodrell Bank Observatory, The University of Manchester, Macclesfield,
Cheshire SK11 9DL, UK}
\altaffiltext{5}
{Auckland Observatory, Auckland, New Zealand}
\altaffiltext{6}
{Department of Astronomy, Ohio State University,
140 W.\ 18th Ave., Columbus, OH 43210, USA; gould@astronomy.ohio-state.edu}
\altaffiltext{7}
{Department of Astronomy, California Institute of Technology, 
Pasadena, CA 91025, USA}
\altaffiltext{8}
{Harvard-Smithsonian Center for Astrophysics, Cambridge, MA 02138, USA}
\altaffiltext{9}
{Department of Physics, Institute for Basic Science Research,
Chungbuk National University, Chongju 361-763, Korea}
\altaffiltext{10}
{American Museum of Natural History, New York, NY}
\altaffiltext{11}
{Farm Cove Observatory, Centre for Backyard Astrophysics,
Pakuranga, Auckland New Zealand}
\altaffiltext{12}
{Bohyunsan Optical Astronomy Observatory, Korea Astronomy and
Space Science Institute, Youngchon 770-820, Korea}
\altaffiltext{13}
{Department of Physics, Notre Dame University, Notre Dame, IN 46556, USA}
\altaffiltext{14}
{Institute for Information and Mathematical Sciences, Massey University,
Auckland, New Zealand}
\altaffiltext{15}
{Solar-Terrestrial Environment Laboratory, Nagoya University,
Nagoya 464-8601, Japan}
\altaffiltext{16}
{Department of Physics, University of Auckland, Auckland, New Zealand;
p.yock@auckland.ac.nz}
\altaffiltext{17}
{Institut d'Astrophysique de Paris, 98bis Boulevard Arago, 75014 Paris, France;
beaulieu@iap.fr}
\altaffiltext{18}
{SUPA St.Andrews, Physics \& Astronomy, St.Andrews, KY16~9SS, UK}
\altaffiltext{19}
{Astrophysics Research Institute, Liverpool John Moores University,
Twelve Quays House, Egerton Wharf, Birkenhead CH41 1LD, UK}
\altaffiltext{20}
{University of Tasmania, School of Maths Physics, Private bag 37, Hobart, 
Tasmania, Australia}
\altaffiltext{21}
{Universit\"at Potsdam, Astrophysik, Am Neuen Palais 10, D-14469 Potsdam,
Germany}

\begin{abstract}
We report the discovery of a several-Jupiter mass planetary companion to
the primary lens star in microlensing event OGLE-2005-BLG-071.  Precise
($\la 1\%$) photometry at the peak of the event yields an extremely high
signal-to-noise ratio detection of a deviation from the light curve
expected from an isolated lens.  The planetary character of this
deviation is easily and unambiguously discernible from the gross
features of the light curve. Detailed modeling yields a
tightly-constrained planet-star mass ratio of $q=m_p/M=0.0071\pm0.0003$.
This is the second robust detection of a planet with microlensing,
demonstrating that the technique itself is viable and that planets are
not rare in the systems probed by microlensing, which typically lie
several kpc toward the Galactic center.  
\end{abstract}

\keywords{gravitational lensing -- planetary systems -- Galaxy: bulge}

\section{Introduction
\label{intro}}

As compared to the other three methods that have successfully detected
extrasolar planets, microlensing has unique features, both positive and
negative.  Unlike pulsar timing \citep{pulsar}, radial velocities
(\citealt{mayor04}; \citealt{marcy05}, and references therein), and
transits \citep{udalski02,konacki03,bouchy04,alonso04},  which rely on
the detection of photons from the host star and thus are biased toward
nearby systems, microlensing is sensitive to mass, and therefore can
detect planets many kpc from the Sun.  Microlensing is potentially very
sensitive to extremely low-mass (e.g. Mars-like) planets because, in
contrast to all other methods, the strength of the signal  can be quite
large ($\ga 10\%$), and the signal-to-noise ratio (S/N) falls with the
planet mass only  as $m_p^{1/2}$.  Microlensing is unique in its ability
to detect wide-separation planets with periods that exceed the duration
of the experiment.

Microlensing, like most other indirect methods, is primarily  sensitive
to the planet-star mass ratio $q=m_p/M$. However,  the microlensing
planet host stars are distant and superimposed on a background source
star, so the brightness and mass of the host star are often only weakly
constrained. Just as the $m_p\sin i$  ambiguity for radial-velocity
planet detections can be broken if the planet happens to transit its
host, so microlensing can yield star and planet masses in special
circumstances as well \citep{bennett02,gould03}. Microlensing detections
occur only at a single epoch, so one measures only the planet-star
separation at a particular moment, not the orbit.

To date, there has been only one robust detection of a planet using the
microlensing technique \citep{ob03235}.  Here we report on the  second
detection of a planet by microlensing, which was enabled by the rapid
response of a number of observing teams to  high-magnification
(high-mag) microlensing events, which are  intrinsically the most
sensitive to planets.  This detection demonstrates that the microlensing
method has reached maturity, and that the planets to which the method is
sensitive are not rare.

\section{High-Magnification Planetary Microlensing Events
\label{sec:highmag}}

When a stellar lens in a microlensing event has a planetary companion,
the binary nature of the lens results in caustics (closed curves of
formally infinite magnification) in the magnification pattern.  If the
source passes over or close to one of these caustics, the light curve
exhibits a short-lived deviation from the standard \citet{pac86}
single-lens form that betrays the presence of the planet \citep{mao91}. 
The great majority of planetary companions generate two distinct types
of caustics: one ``central caustic'' that lies close to the host star
and one or two ``planetary caustics'' depending on whether the planet
lies outside or inside the Einstein ring. As one moves the planet closer
to the Einstein ring, both types of caustics grow, and they eventually
merge into a single ``resonant caustic''.

\citet{gouldloeb92} pointed out that planets in Jupiter-like orbits
around stars on the line of sight  to the Galactic bulge would
coincidentally lie fairly close to the Einstein ring. Hence, they should
generate large and so easily detectable planetary caustics.  This
optimistic assessment encouraged early efforts to detect planets.
Because planetary caustics are much larger than central caustics, in an
unbiased sample of microlensing events the overwhelming majority of
planetary anomalies will be generated by planetary caustics, and for
this reason they were the focus of early efforts.

Nevertheless, central caustics play a crucial role in current
microlensing planet searches.  Exactly because they are ``central'',
central caustics can be probed only in very  high-mag events, i.e.,
events in which the source passes very close to the primary lens and so
to the central caustic.  These events are very rare, but when they do
occur, the source is essentially {\it guaranteed}  to  pass over or
close to the central caustic if a planet is present 
\citep{griestsafi,bond02,ratten02}. Since very high-mag events can be
identified as they are unfolding, they can be intensively monitored over
their peak to a degree  that is not feasible for the much more frequent
garden-variety microlensing events.   As a result, modern microlensing
followup groups like PLANET (\citealt{planet}) and $\mu$FUN
(\citealt{yoo}) now tend to focus much of their effort on identifying
and following the rare high-mag events.  Moreover, both of the major
survey teams, OGLE (\citealt{udalski93}) and MOA (\citealt{moa}), switch
over from survey mode to followup mode when events undergo high-mag or
other effects that warrant closer monitoring.

The problem of identifying high-mag events is actually  quite severe.  
The great majority of high-mag sources are faint, and as a result the
photometry of the early light curve is generally too poor to accurately
predict their high-mag character well in advance.   As a result, fewer
than a dozen high-mag $(A_\max\ga 100)$ events  have been intensively
monitored altogether \citep{five,gaudi02,rhie00,moa,science,yoo,dong05}.

The OGLE-III Early Warning System (EWS, \citealt{ews}) now annually
alerts bulge microlensing events at the rate of 600 per year.  This
provides a potentially rich source of high-mag events.  In order to
better harvest this potential, as well as to react quickly to
microlensing anomalies, OGLE has now implemented the  Early Early
Warning System (EEWS), which detects anomalies in real time.

\section{Event Recognition and Observational Data
\label{data}}

The original alert on OGLE-2005-BLG-071 was triggered by EWS \citep{ews}
on 17 March 2005 based on observations by OGLE-III.  The alert predicted
that the event would peak about 1 month later on HJD$'\equiv
$HJD$-2450000=3880\pm 8$, with $A_\max > 3$.  Based on a new OGLE point
(HJD$'$ 3477.9) less than 3 days before the lens-source point of closest
approach, as well as its own data (acquired starting HJD$'$ 3472.7),
$\mu$FUN issued a general alert (HJD$'$ 3478.20) that the event was
peaking at high magnification. This triggered more intensive
observations the next night by OGLE and by $\mu$FUN Chile.

A single-lens fit to these new observations resulted in substantially
worse $\chi^2$ than previous fits, and so $\mu$FUN then issued a second
alert (HJD$'$ 3478.97) saying that an anomaly had begun.  This was
confirmed with the first OGLE observation on the following night (HJD$'$
3479.73). At this point, both OGLE and all $\mu$FUN stations attempted
to observe the event intensively and continued to do so for the next 4
nights. OGLE was able to obtain data from Chile during almost the entire
period and indeed issued an EEWS alert on HJD$'$ 3481.94 reporting on a
second rise.  Two New Zealand $\mu$FUN observatories (Farm Cove and
Auckland) obtained substantial data including 6 continuous hours on the
falling sides of each of the ``twin peaks'' of the light curve, and
$\mu$FUN observations from Palomar observatory yielded coverage of the
second peak, thereby bridging the rise and fall covered by OGLE and
Farm-Cove/Auckland respectively.  A small amount of additional  data
covering the second peak was obtained from Kitt Peak.  MOA immediately
responded to the original high-mag alert to obtain data on the rise
toward the first peak as well as all four nights of the anomaly. 
PLANET/RoboNet, which had already made pre-peak observations beginning
HJD$'$ 3470.1, responded to the anomaly alert, catching the decline of
the first peak from the Faulkes North Telescope in Hawaii and the second
from the Canopus Telescope (Tasmania).

It was soon realized that the anomaly was short-lived and on HJD$'$
3482.9 the light curve returned to the normal single-microlens shape.
The first results of modeling of the light curve of the event announced
on HJD$'$ 3483.9 by OGLE suggested the possibility that the anomaly was
due to a low mass companion, well into the planetary regime. This result
was subsequently confirmed by independent modeling conducted by other
teams.

We present data from OGLE (1.3m at Las Campanas Observatory in Chile,
operated by the Carnegie Institution of Washington), MOA (0.6m telescope
at Mt. John University Observatory in New Zealand) $\mu$FUN Chile
(SMARTS 1.3m telescope at CTIO), Palomar ($60''$ Robotic telescope), MDM
(Hiltner 2.4m at Kitt Peak), Auckland (0.35m Nustrini telescope at
Auckland Observatory), Farm Cove (0.25m Meade telescope at Farm Cove
Observatory), Faulkes North (2.0m in Hawaii) and Canopus (1.0m at
Hobart, Tasmania).  

\begin{figure}
\scalebox{0.44}[0.44]{\includegraphics{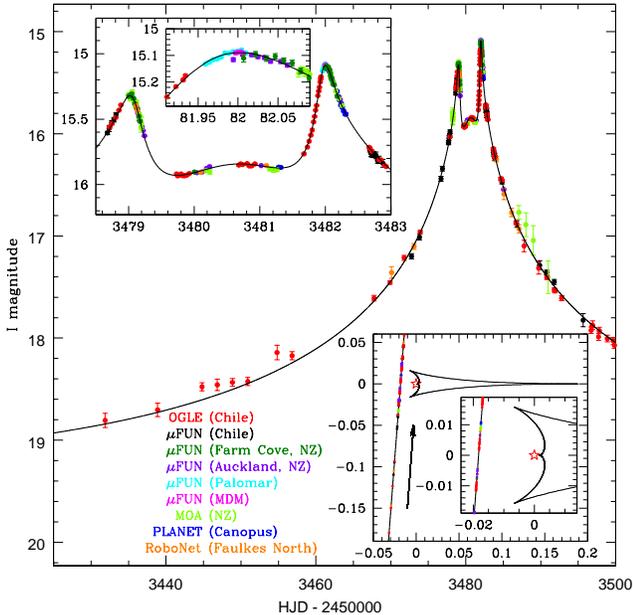}}
\caption{Inspection of the light curve of microlensing event
OGLE-2005-BLG-071 shows it contains a planet.   Main figure shows that,
apart from the anomaly near the peak, this was an ordinary high-mag
event, implying that the caustic is small.  The triple peak (two big
symmetric peaks surrounding a small peak) shows that the source passed 3
cusps of a caustic, with the middle one being weak (see inset), which
implies a normalized star-companion separation $b\sim 1$.  Interval
between peaks (and so between cusps) is $\Delta t=3\,$days, implying the
shear induced by the companion is small, $\gamma = \Delta t/4t_\e <
0.02$, so the mass ratio $q$ of the companion is also small, $q= \gamma
b^2 \la 0.03$.  More detailed fitting shows $q=0.0071$.
}
\label{fig:lightcurve}
\end{figure}

\section{Only a Planet Can Explain This Light Curve\label{sec:modeling}}

Figure \ref{fig:lightcurve} shows the data from the various
observatories together with a binary-lens  model.  The model shown has a
very small mass ratio, $q=0.0071$, and is clearly a good fit to the
data. The event is therefore consistent with a planetary lens, however
the question remains whether there might be other non-planetary models
that fit the data equally well, i.e.\ binary-lens models with $q\sim
O(1)$.   Binary lens models have 7 lens parameters plus $2n$ flux
parameters where $n=10$ is the number of observatory/filter
combinations.  Three parameters are the same as for single lenses, the
time of closest approach to the lens ``center'' $t_0$, the impact
parameter (normalized to the angular Einstein radius $\theta_\e$) $u_0$,
and the Einstein radius crossing time $t_\e$.  Three parameters specify
the lens geometry, the mass ratio $q$, the separation of the two
components (normalized to $\theta_\e$) $b$, and the angle of source-lens
relative motion with respect to the binary axis $\alpha$.  In addition,
if the source is resolved by the magnification pattern, one must specify
the ratio of the source size to the Einstein radius,
$\rho=\theta_*/\theta_\e$. Finally, for each observatory/filter
combination, there is a source flux $f_s$ and a background flux $f_b$
such that the total flux is $f=A f_s + f_b$ where $A$ is the
magnification.

Given that the parameter space is obviously very large and somewhat
complex, how do we know that there are no non-planetary solutions? One
way to tell is to conduct a wide search for solutions, which we have
done.  However, it is also useful to have analytic arguments to ensure
that a solution is not lurking in a corner of parameter space that one
did not try.

The following argument rests just on the gross features of the light
curve.  First, the anomaly occurs near the peak of an otherwise normal
event, when it is 3 magnitudes above baseline, so $A>16$ i.e., the
normalized source-lens separation is $u<0.06$.  This already implies
that the caustic is small, and so must be either a central caustic
(generated by a wide or close companion) or the ``central-caustic end''
of a resonant caustic. The central caustics of wide and close binaries
(with $b\leftrightarrow  b^{-1}$) are mathematically nearly identical
\citep{dominik99,jin05}, and the central-caustic end of a resonant
caustic is very similar to these.  The twin-peaked structure can only be
generated by the source passing close to, but not over, two cusps of
this central caustic.  Peaks can also be caused by passing over a
caustic, but in that case they are highly asymmetric, with a much faster
rise for the first peak and a much faster decline for the second.  This
alternate scenario is clearly ruled out by the form of the light curve. 
Third there is a small ``bump'' in between these two peaks.  This can
only be caused by passing a third, much weaker cusp.  All of these
features are matched by the caustic geometry shown in the inset and
subinset to Figure~\ref{fig:lightcurve} and cannot be matched by
caustics that lack this three-pronged morphology.

The fact that the two peaks are of almost equal height implies that the
source must pass nearly perpendicular to the binary axis, $\alpha\sim\pm
90^\circ$. The fact that the middle peak is so much weaker than the
outer two implies that the caustic is extremely asymmetric.  Such
asymmetric caustics occur only when $b$ is close to unity: for $b\gg 1$
or $b\ll 1$, the caustics are diamond-shaped. For definiteness, we now
consider the wide-binary case $b>1$.  Since the central caustic is
similar to a \citet{cr1} caustic, its full width is equal to $4\gamma$,
where $\gamma = q/b^2$ is the shear, so the time to cross between these
cusps is $\Delta t= 4q t_\e/b^2$. Equating this to the time between the
peaks, $\Delta t\sim 3\,$days, yields $q = b^2(0.75 \,{\rm days}/t_\e)$.
 Since the blending is known from the full  fit to the light curve,
$t_\e$ is well determined from the fit to the light curve away from the
peak.  However, even ignoring this information, the maximum $q$ can be
found from the minimum allowed $t_\e$, which is obtained by assuming no
blending. This minimum is $t_\e\sim 40\,$days.  That is, $q< 0.019b^2$
or $q\la 0.03$.  This limit is already very close to the planet regime. 
Moreover, the accumulated constraints on $b$, $q$, and $\alpha$ imply
that the allowed parameter space is small, and so was easily and
exhaustively searched. A virtually identical argument applies to the
$b<1$ case, yielding a planet of virtually the same mass.   The best-fit
parameters  for both wide and close solutions are given in 
Table~\ref{tab:pars}.

\begin{deluxetable*}{
c@{\hspace{5pt}}
c@{\hspace{5pt}}
c@{\hspace{5pt}}
c@{\hspace{5pt}}
c@{\hspace{5pt}}
c@{\hspace{5pt}}
c@{\hspace{0pt}}
c@{\hspace{0pt}}
c@{\hspace{0pt}}
c@{\hspace{5pt}}
}
\tablecaption{\sc OGLE-2005-BLG-071 Model Parameters}
\tablewidth{0pt}
\tabletypesize{\scriptsize}
\tablehead{
  \colhead{Model} &
  \colhead{$t_0$} &
  \colhead{$u_0$} &
  \colhead{$t_\e$} &
  \colhead{$q$} &
  \colhead{$b$} &
  \colhead{$\alpha$} &
  \colhead{$I_s$} &
  \colhead{$I_b$} &
  \colhead{$\chi^2$}\\
  \colhead{} &
  \colhead{(HJD-2453400.)} &
  \colhead{} &
  \colhead{(days)} &
  \colhead{} &
  \colhead{} &
  \colhead{($^\circ$)} &
  \colhead{(mag)} &
  \colhead{(mag)} &
  \colhead{(1092 dof)}
}
\startdata
Wide & 80.6791$\pm$0.0020  & 0.0236$\pm$0.0013 & 70.9$\pm$3.3 &  0.0071$\pm$0.0003 & 1.294$\pm$0.002 & $274.23\pm 0.04$ & 19.53 & 21.29 & 1105.6\\
Close & 80.6919$\pm$0.0023  & 0.0225$\pm$0.0012 & 73.9$\pm$3.5 &  0.0067$\pm$0.0003 & 0.758$\pm$0.001 & $274.48\pm 0.05$ & 19.59 & 21.05 & 1127.6\\
\enddata
\label{tab:pars}
\end{deluxetable*}

The event is still in its late phases. When it reaches baseline a more
detailed analysis of the light curve may permit measurements of
additional parameters, in particular the mass of the lens star. Based on
a preliminary analysis of finite-source effects during the second peak
and parallax effects in the wings of the event (c.f. \citealt{an02}) we
constrain the host mass to be $0.08 < M/M_\odot < 0.5$, implying that
the planet lies in the range $0.05<m_p/M_{\rm jup}<4$.  The
corresponding range of distances is $1.5<(D_l/{\rm kpc})<5$. 

We obtain an upper limit on the lens-star flux, $I_l > 21.3$, by
analyzing good-seeing OGLE images constrained by astrometry derived from
a {\it Hubble Space Telescope (HST)} image taken on HJD$'$ 3513.6. Our
estimate of the mass/distance range is fully consistent with this limit
implying that the lens star cannot be heavier.  When a second {\it HST}
image is taken after the event, the flux from the lens primary will be
constrained even more precisely.

\section{Discussion
\label{sec:discuss}}

The discovery of a planet in OGLE-2005-BLG-071 has several important
implications.  First, being the second such detection, it shows that
microlensing planets are not a fluke.  While the Poisson statistics of a
single detection were consistent with extremely low rates, two
detections makes the low-rate hypothesis implausible.

Second, OGLE-2005-BLG-071 is the first secure high-mag planetary event. 
While \citet{griestsafi} long ago identified these  central-caustic
probing events as a potentially rich vein for planet hunting, the
technical difficulties in recognizing high-mag event in real time and in
adequately monitoring their peak have limited their application to a few
events.  With the coming on line of  OGLE EEWS and the increasing
sophistication of followup teams, excellent coverage of high-mag events
is becoming more common, if not quite routine. 

Finally, in addition to being a fruitful path to planets in general,  
central-caustic events are at present the only practical method for
finding  Earth-mass planets around main-sequence stars using current
technology.   Although OGLE-2005-BLG-071 contains a giant planet, the
fact that it was detected at extremely high S/N demonstrates that much
lower-mass planets can also be detected.

\acknowledgments

Support for OGLE was provided by Polish MNII grants 2P03D02124,
2P03D01624, NSF grant AST-0204908 and NASA grant NAG5-12212. A.U.\ was
supported from the grant SP13/2003 of the Foundation for Polish Science.
A.G.\ and S.D.\ were supported by NSF grant AST 02-01266. D.D., A.G.,
and R.P.\ were supported by NASA Grant NNG04GL51G. A.G.-Y.\ was
supported by NASA through a Hubble Fellowship. B.S.G.\ was supported by
a Menzel Fellowship from the Harvard College Observatory.   C.H.\ was
supported by the SRC program of Korea Science \& Engineering Foundation.
B.-G.P.\ acknowledges support from the Korea Astronomy and Space Science
Institute. MOA thanks the Univ.\ of Canterbury for making
available the Mt John Observatory 0.6m telescope. Work by MOA is
supported by the Marsden Fund of New Zealand and the Ministry of
Education, Culture, Sports, Science and Technology of Japan.  The Mt
Canopus Observatory is operated by the University of Tasmania with
financial assistance from David Warren. RoboNet is a PPARC-funded
project led by Liverpool John Moores University on behalf of a
consortium of 10 UK universities.  The Faulkes Telescope North is
operated with support from the Dill Faulkes Educational~Trust.


\begin{thebibliography}{99}

\bibitem[Abe et al.(2004)]{science}
Abe, F. et al.\ 2004, Science, 305, 1264

\bibitem[Albrow et al.(1998)]{planet}
Albrow, M.D. et al.\ 1998, \apj, 509, 687

\bibitem[Albrow et al.(2001)]{five}
Albrow, M.D. et al.\ 2001, \apj, 556, L113

\bibitem[Alonso et al.(2004)]{alonso04} 
Alonso, R., et al.\ 2004, \apjl, 613, L153 

\bibitem[An et al.(2002)]{an02} An, J.H. et al. 2002, \apj, 572, 521

\bibitem[An(2005)]{jin05} An, J.H.  2005, \mnras, 356, 1409

\bibitem[Bennett \& Rhie(2002)]{bennett02} 
Bennett, D.P., \& Rhie, S.H.\ 2002, \apj, 574, 985

\bibitem[Bond et al.(2001)]{moa}
Bond, I. A., et al.\ 2001, \mnras, 327, 868

\bibitem[Bond et al.(2002)]{bond02}
Bond, I. A., et al.\ 2002, \mnras, 331, L19

\bibitem[Bond et al.(2004)]{ob03235}
Bond, I.A., et al.\ 2004, \apj, 606, L155

\bibitem[Bouchy et al.(2004)]{bouchy04} 
Bouchy, F., Pont, F., Santos, N.~C., Melo, C., Mayor, M., 
Queloz, D., \& Udry, S.\ 2004, \aap, 421, L13 

\bibitem[Chang-Refsdal(1979)]{cr1} 
Chang, K.\ \& Refsdal, S.\ 1979, Nature, 282, 561

\bibitem[Dominik(1999)]{dominik99} 
Dominik, M.\ 1999, \aap, 349, 108

\bibitem[Dong et al.(2005)]{dong05} Dong, S. et al. 2005, in preparation

\bibitem[Gaudi et al.(2002)]{gaudi02} 
Gaudi, B. S., et al.\ 2002, \apj, 566, 463

\bibitem[Gould et al.(2003)]{gould03} 
Gould, A., Gaudi, B.S., \& Han, C. 2003, \apj, 591, L53

\bibitem[Gould \& Loeb(1992)]{gouldloeb92} 
Gould, A., \& Loeb, A.\ 1992, \apj, 396, 104

\bibitem[Griest \& Safizadeh(1998)]{griestsafi}
Griest, K.\ \& Safizadeh, N.\ 1998, \apj, 500, 37

\bibitem[Konacki et~al.(2003)]{konacki03} 
Konacki, M., Torres, G., Jha, S., \& Sasselov, D.~D.\ 2003, \nat, 421, 507

\bibitem[Mao \& Paczy\'nski(1991)]{mao91}
Mao, S.\ \& Paczy\'nski, B.\ 1991, \apj, 374, 37

\bibitem[Marcy et al.(2005)]{marcy05} Marcy, G.W., Butler, R.P., Vogt, S.S., 
Fischer, D.A., Henry, G.W., Laughlin, G., Wright, J.T., \& Johnson, J.A. 
2005, \apj, 619, 570

\bibitem[Mayor et al.(2004)]{mayor04} Mayor, M., Udry, S., 
Naef, D., Pepe, F., Queloz, D., Santos, N.~C., \& Burnet, M.\ 2004, \aap, 
415, 391 

\bibitem[Paczy\'nski(1986)]{pac86}
Paczy\'nski, B.\ 1986, \apj, 304, 1

\bibitem[Rattenbury et al.(2002)]{ratten02}
Rattenbury, N.J., Bond, I.A., Skuljan, J., \& Yock, P.C.M, 2002, \mnras, 335, 159

\bibitem[Rhie et al.(2000)]{rhie00}
Rhie, S. H. et al. 2000, \apj, 533, 378

\bibitem[Udalski et al.(2002)]{udalski02} 
Udalski, A., et al.\ 2002, Acta Astron., 52, 1 

\bibitem[Udalski(2003)]{ews}
Udalski, A. 2003, Acta Astron., 53, 291

\bibitem[Udalski et al.(1993)]{udalski93} Udalski, A., Szyma\'nski, M., 
Ka{\l}u\.{z}ny, J., Kubiak, M., Krzemi\'nski, W., Mateo, M., Preston, G.W., 
Paczy\'nski, B. 1993, Acta Astron., 43, 289 

\bibitem[Wolszczan \& Frail(1992)]{pulsar} Wolszczan, A. \& Frail, D.A. 1992,
Nature, 355, 145

\bibitem[Yoo et al.(2004)]{yoo} Yoo, J., et al.\ 2004, \apj, 616, 1204

\end{thebibliography}
\end{document}